%% file: main.tex
\documentclass[conference]{IEEEtran}
\usepackage[utf8]{inputenc}
\usepackage{lipsum}

\usepackage{subcaption}
\usepackage[frozencache]{minted}

\usepackage{url}
\usepackage{xspace}
\usepackage{graphicx}
\usepackage{booktabs}

\usepackage{enumitem}
\usepackage{tikz}

\usepackage[para,online,flushleft]{threeparttable}
\usepackage{tabulary}
\usepackage{hyperref}

\input{macros.tex}

\begin{document}

\title{\TOOL: Evaluating the Executability of \\Python Code Snippets on GitHub}

\author{
    \IEEEauthorblockN{Eric Horton, Chris Parnin}
    \IEEEauthorblockA{
        NC State University\\
        Raleigh, NC, USA\\
        Email: \{ewhorton, cjparnin\}@ncsu.edu
    }
}

\maketitle

\begin{abstract}

Software developers create and share code online to demonstrate programming language concepts and programming tasks. Code snippets can be a useful way to explain and demonstrate a programming concept, but may not always be directly executable. A code snippet can contain parse errors, or fail to execute if the environment contains unmet dependencies.

This paper presents an empirical analysis of the executable status of Python code snippets shared through the GitHub gist system, and the ability of developers familiar with software configuration to correctly configure and run them. We find that 75.6\% of gists require non-trivial configuration to overcome missing dependencies, configuration files, reliance on a specific operating system, or some other environment configuration. Our study also suggests the natural assumption developers make about resource names when resolving configuration errors is correct less than half the time.

We also present \TOOL, a database and extensible framework built on GitHub's gist system, which provides executable code snippets to enable reproducible studies in software engineering. \TOOL contains 10,259 code snippets, approximately 5,000 with a Dockerfile to configure and execute them without import error. \TOOL is publicly available at \url{https://github.com/gistable/gistable}.


\end{abstract}

\section{Introduction}

Online programming communities such as Stack Overflow and GitHub facilitate social learning of programming and API concepts. One common learning mechanism is to share code snippets or examples, which contain explanations and demonstrate how to perform a programming task or use an API~\cite{Sillito:2012}. Code snippets are often reused and incorporated in open source projects~\cite{Yang:2017}. Currently, GitHub provides the ability to create and share code snippets (called \emph{gists}), with over 300k Python gists, and over 4.5 million gists in multiple programming languages.

This work focuses on evaluating the excitability of publicly available Python scripts hosted on GitHub's gist system in the context of software configuration management (the process of configuring system environments to properly execute a software program). We seek to categorize the common reasons for why gists cannot be executed in a default environment and motivate further research on automated software configuration management by identifying what difficulties exist in properly configuring an environment to enable gist execution, specifically with regard to installing application dependencies. Our work also provides a dataset of gists known to not be executable and a baseline analysis against which to compare future configuration methods.

We start by highlighting a method for performing automated gist collection and analysis. The process involves scraping gist URLs from the GitHub gist UI. This technique led to an initial dataset of 10,259 gists containing over 1,700 unique third-party library packages. We then cloned each gist and executed it inside of a Docker container based on the official Python image for Docker, categorizing the gist by its exit status. To evaluate gist configuration, we attempted to infer a correct environment specification using a naive algorithm that approximates the first steps human developers often take by attempting to install third-party packages by the name of the resource imported within the gist. After running the naive inference algorithm, we then reevaluated the executable status of the gist.

Our findings show that correct dependency resolution and environment configuration are often required even for small programs. Less than 25\% of gists were executable by default, with over half failing due to \texttt{ImportError} in Python 2. Of the gists which initially failed with \texttt{ImportError}, our naive inference algorithm could successfully infer an environment specification less than 50\% of the time.

To gain a better understanding of the why the naive inference algorithm fails, we asked 24 developers familiar with system configuration practices to create a Dockerfile for 10 unique gists, assigned to them at random, for which inference failed to resolve import errors. We then used the produced Dockerfiles and feedback from the developers to categorize gists, focusing on the first cause of failure if any gist would have failed inference due to more than one reason. The most common cause is that the names of resources used in a gist do not necessarily match the names of the packages they belong to. Gists also frequently fail due to missing transitive dependencies, missing system dependencies, configuration files, and deprecated or non-standard packages.

Finally, we present \TOOL{}, an extensible database and framework used to perform our mining and analysis. \TOOL{} also contains the \textasciitilde{}5k gists with environment specifications which allow them to be run without \texttt{ImportError}. We believe that several areas of software engineering research can benefit from a database of executable code snippets, such as: automatic code summarization, testing, and API usage analysis.

In summary, this paper makes the following contributions:

\begin{itemize}
    \item An empirical analysis on the executable status of Python gists on GitHub.
    \item A qualitative analysis of reasons why code may not be executable.
    \item An extensible mining framework, \TOOL, for obtaining gists and environment containers from our gist database.
\end{itemize}

\section{Motivation}

Code snippets are not always directly usable~\cite{Yang:2016}: They can contain parse errors or require system dependencies unmet in a programming environment. As a result, the following challenge emerges: \emph{Given a code snippet, successfully infer the environment configuration necessary for execution.} Frequently, developers must perform this inference step manually, or rely on the creation of configuration scripts, which in itself is a time consuming task~\cite{McIntosh:2011}. Unfortunately, it is not always clear what dependencies or environment configurations are required to execute code. Consider the following Python code snippet.

\begin{minted}[fontsize=\footnotesize,linenos,xleftmargin=0.5cm,breaklines,autogobble]{python}
    # Import modules from networkx and matplotlib
    from networkx.drawing.nx_agraph import graphviz_layout
    import matplotlib.pyplot as plot
    import networkx as nx

    # Generate the complete graph on five vertices
    k5 = nx.complete_graph(5)

    # Draw using layout generated by graphviz
    plot.figure()
    nx.draw(k5, graphviz_layout(k5, prog="neato"))
    plot.savefig('/output/graph.png')
    plot.close()
\end{minted}

To successfully run this code fragment, several requirements must be met. First, the environment requires graphviz, which is a tool for visualizing graphs. Second, the environment needs to install the Python bindings for graphviz. Third, the environment needs the Python package for matplotlib and networkx. Fourth, an environment variable, \texttt{MPLBACKEND}, may be needed to specify a rendering engine that is compatible with a headless VM, which does not have a graphics display. Finally, the environment needs to ensure that an \texttt{/output} directory exists.

These requirements can also be encapsulated by a working environment configuration. One system that can be used for specifying environment configuration is the containerization system Docker. Docker configuration is centered around the Dockerfile, a configuration script which tells the Docker engine how to properly build an image that can be distributed and run by others. We present a Dockerfile for the snippet below.

\begin{minted}
[
fontsize=\footnotesize,linenos,xleftmargin=0.5cm
%, style=borland
]
{Dockerfile}
FROM python:2.7.13
VOLUME /output
ENV MPLBACKEND Agg
RUN apt-get update
RUN apt-get install -y graphviz
RUN pip install pygraphviz
RUN pip install matplotlib
RUN pip install networkx
ADD snippet.py /scripts/
CMD python /scripts/snippet.py
\end{minted}

\section{\TOOL Dataset and Tool}

\begin{figure*}[ht!]
    \centering
    \begin{tabular}{c|c}
        \begin{subfigure}[b]{0.55\textwidth}
            \centering

\begin{minted}[fontsize=\footnotesize,linenos,autogobble,breaklines,breakafter=/]{Python}
    import requests
    import json

    urlbase = 'http://maps.googleapis.com/maps/api/geocode/json?sensor=false&address='
    urlend = 'Zurich,Switzerland'

    r = requests.get(urlbase+urlend) # request to google maps api

    r=r.json()
    if r.get('results'):
        for results in r.get('results'):
            latlong = results.get('geometry','').get('location','')
            latitude = latlong.get('lat','')
            longitude = latlong.get('lng','')
            break
        print latitude, longitude

    else:
        print 'No results'
\end{minted}
\caption{\href{https://gist.github.com/philshem/10017416}{Gist 10017416}}
\label{fig:motivation:code}
\end{subfigure}
&
\begin{subfigure}[b]{0.35\textwidth}
\centering
\begin{minipage}{.8\textwidth}
\begin{minted}[fontsize=\footnotesize,linenos,autogobble]{Dockerfile}
    FROM python:2.7.13
    ADD snippet.py snippet.py
    RUN ["pip", "install", "requests"]
    CMD ["python", "snippet.py"]
\end{minted}
\end{minipage}
\caption{Dockerfile}
\label{fig:motivation:Dockerfile}
\end{subfigure}
\end{tabular}
\caption{(a) Code snippet for using the Google Maps geocode API. (b) Dockerfile containing environment specification required to run code snippet. }
    \label{fig:gistable}
\end{figure*}

Gistable is a framework for collecting, evaluating and executing self-contained programming code snippets, called gists. The name is derived from a portmanteau of the words \emph{gists} and \emph{runnable}. Gistable is designed to support empirical research for a variety of software engineering tasks. \TOOL can mine code snippets and automatically generate a Dockerfile which can be used to run the code snippet. \TOOL provides a command line interface for performing tasks with the mined gists, such as checking out snippets into a working directory, and executing the code snippet inside a docker container.

\subsection{Research context.}

Our initial evaluation of \TOOL focuses on Python gists. Python is a popular programming language and ranks among the fastest growing languages today. It follows only Ruby and Javascript in proportion of files in public gists~\cite{Wang:2015:GUU:2820518.2820556}. Python is frequently used for teaching introductory programming classes as well as used by non-professional programmers, such as scientists.

Previous research by Yang et al.~\cite{Yang:2016} examined Python snippets on Stack Overflow and found that only 25\% were runnable (but did not investigate why). In this work, we focus on examining gists shared on GitHub instead of Stack Overflow. As observed by Sillito et al. ~\cite{Sillito:2012} and Yang et al.~\cite{Yang:2016}, code snippets on Stack Overflow are often mixed with exposition and code, making it difficult to understand which segments of code are meant to be executed in an automated analysis. Therefore, there is strong motivation to investigate the underlying reasons why Python code may not be executable and understand the effort involved in configuring environments capable of running it. These barriers can cause problems for learners and non-professionals programmers lacking system configuration skills.

\subsection{Mining Gists}\label{subsec:mining-gists}

We consider two strategies for mining gists from GitHub. GitHub provides a REST API for public gists, however, there are several limitations. Currently, the API provides no support for filtering queries based on language type. Furthermore, the API limits requests to 3000 gists when using pagination. To overcome these limitations, it is possible to filter gists based on creation date, meaning that all gists could be slowly enumerated by strategically modifying the creation date as a filter.

Another strategy is to scrape gists from the GitHub gist search UI. The search UI  allows several filters, such as star rating, language, and keywords contained in the gist. The UI returns at most 100 pages of 10 random gists matching a search, allowing 1000 gists to be returned per search. By strategically modifying search terms, it is possible to quickly discover gists that meet the desired criteria.

For our initial population of the \TOOL database, we focused on the scraping approach, which allowed us to focus on a particular language and to better control the quality of gists while using less computational resources.


\subsection{Environment Inference Algorithm}\label{sec:inference}

To perform environment inference, we use an approach which builds an Abstract Syntax Tree (AST) of the gist source code and extracts all declared imports. Extracted imports are then filtered to remove all packages which are part of the Python standard library. Imports are assumed to be part of the standard library if they are present in a Docker image containing a clean install of the Python runtime.

We use the assumption that each import represents a single package that needs to be installed, and that the import name matches the name used to install the package. This is not always the case. For example, the Python package \texttt{beautifulsoup4} is imported as \texttt{bs4}. However, developer practices from Section~\ref{subsubsec:rqwnenv-description} suggest that this is a useful approximation because it is the natural first step a developer takes when attempting to configure a computing environment. Errors from packages which could not be found are ignored. Such packages are simply not included in the final environment configuration. This allows us to recover from potential errors in our inference algorithm.

\subsection{Execution Harness}\label{sec:harness}
To deal with the large number of gists analyzed as part of the \TOOL database, we built an execution harness on a distributed cluster using the HashiCorp Nomad job scheduler, which natively supports docker containers. The harness is responsible for running all gists through the validation process to first determine if environment inference is needed and categorize the result of gist execution.

To isolate effects of dependencies and other system wide configurations, we perform analysis inside independent Docker containers. The container filesystem also guarantees consistent starting environments.

\subsection{Using \TOOL}

\TOOL provides a command line tool for interacting with gists from the \TOOL database. Gists can be cloned into a specified directory using the command \texttt{gistable clone <id> [location]}. Behavior is similar to that of \texttt{git clone}, and gists are checked out to the working directory if no location is specified.

If Docker is installed and running on the system, the CLI can also be used to directly execute a gist and display all execution results. Just call \texttt{gistable run <id>}.

\section{Methodology}

\subsection{Research Questions}

In this study, we investigate the following research questions and offer the motivation for each:

\emph{\RQExec} Can the average Python gist on GitHub be run to completion, or will it raise an exception? If gists can be run to completion, then they already form a database of snippets that can be used in research. However, if, like the Python snippets from \cite{Yang:2017}, gists cannot be executed by default due to syntax errors or other runtime exceptions, then additional investigation is needed.


\emph{\RQEnv} Can we apply a simple approach for resolving unmet Python dependencies to address most runtime exceptions? If a majority of errors can be addressed by a simple resolution strategy, then there are a limited number of cases where automated environment configuration is needed. However, if a simple approach cannot be used, then more research is needed for developing a more comprehensive automatic environment configuration technique.

\emph{\RQWNEnv} If gists cannot be executed even after resolving package dependencies, the natural question is why. Are they missing configuration for environment variables, services, or other kinds of dependencies? Categorizing gist execution failure and finding common root causes may lead to insight into how to improve future automatic environment configuration techniques.

\subsection{Data Collection}

To address our research questions, we first focused on building a large dataset of Python gists. We used the mining procedure outlined in Section~\ref{subsec:mining-gists} to mine 10,259 Python gists. We limited our search criteria to gists with at least one star~\cite{Kalliamvakou:2014}. Currently, GitHub contains 32,233 Python gists with at least star\textemdash meaning our sample represents nearly 31\% of all public starred Python gists.

Figure~\ref{fig:gistable} illustrates an example of a gist in our experimental dataset and its accompanying automatically created Dockerfile. The gist uses the Google Maps geocode api to retrieve the latitude and longtitude coordinates of Zurich, Switzerland. The Dockerfile bases the image off of a Python environment, adds the gist code file, installs \texttt{requests}, and configures the default command to run the gist. Note that the package, \texttt{json}, does not need to be installed as it is a default system package.

\subsection{Analysis}

To answer our research questions, we used the following procedures to analyze our data. The inference harness described in Section~\ref{sec:harness} was used to clone gists from GitHub and perform analysis. Using two \texttt{ubuntu-16-04-x64} worker nodes sized at \texttt{2gb} and running in Digital-Ocean, inference took approximately eight hours to schedule and run all jobs.

\subsubsection{\RQExecShort }

To answer \RQExecShort, we start by performing a baseline analysis of gists by attempting to execute them in isolated Docker containers based on the \texttt{python:2.7.13} and \texttt{python:3.6.5} images. Any gist which executed without error is considered to have exited with the code \texttt{Success}. Any non successful gist is coded by the name of the error which was raised. I.E., \texttt{SyntaxError}, \texttt{ImportError}, \texttt{NameError}, etc.

\subsubsection{\RQEnvShort}







Research from Becker et al.~\cite{Becker:2018} indicates that the practical approach when there are multiple failures is to focus on the first error until it is resolved, then move on. This follows from the observation that first failures are useful because they are informative, need to be fixed, and their resolution may reveal deeper errors that were not apparent before.

To answer \RQEnvShort, we focus on gists where the first encountered failure was an \texttt{ImportError} and ask if we can configure the environment with all necessary dependencies. A naive attempt is made at performing environment inference by applying the inference procedure described in Section~\ref{sec:inference}. We attempt to install each inferred package with the Python package manager \texttt{pip}. This is based on our findings from Section~\ref{subsubsec:rqwnenv-description}, which showed that attempting to install a resource name listed in an import error is often the first step developers take when attempting to fix environment configurations.


After applying our inference algorithm, the gist is then executed a second time with the new environment specification, and the evaluation results recorded under the same criteria as for the baseline.

\subsubsection{\RQWNEnvShort}\label{subsubsec:rqwnenv-description}

We performed a random sampling on failing gists in order to understand why they failed to execute. For this analysis, we performed descriptive coding~\cite{Saldana2009} and composed \emph{memos}~\cite{Birks2008}, which described several reasons for a gist failing to execute. These memos captured interesting events or properties of environments and code snippets to promote depth and credibility, and to frame the information needs of an automated environment configuration technique. That is, they provide a \emph{thick description} to contextualize the findings~\cite{Ponterotto2006}.

We then solicited 24 developers familiar with Docker to manually inspect gists. Each developer was given a disjoint random set of 10 gists and asked to create a Dockerfile that would enable successful execution of the snippet within a standard time period (one and a half weeks). The developers had between 6 months to 5 years of industry experience and familiarity with Python. Further, the developers had been trained in several workshops on configuration management skills, including Ansible and Docker.

We asked the developers to rate the difficulty of creating a Dockerfile and the steps they took to create it. We then performed a qualitative coding exercise over the Dockerfiles and reported steps using closed codes derived from our first qualitative coding.
During the coding process, we employed the technique of \emph{negotiated agreement} as a means to address the reliability of coding~\cite{Campbell2013}. Using this technique, the first and second authors collaboratively code to achieve agreement and to clarify the definitions of the codes; thus, measures such as inter-rater agreement are not applicable.

\section{Executability Results}\label{sec:evaluation}

\subsection{\RQExec}

    \begin{table}
        \caption{
            Gists per exit code in the baseline evaluation using Python 2.7.13.
        }
        \label{tbl:evaluation-before-naive-inference}
        \centering
        \begin{tabular}{l|r|r}
            \textbf{Result} & \textbf{Count} & \textbf{Percent} \\
            \midrule
            ImportError & 5379 & 52.4\% \\
            Success & 2501 & 24.4\% \\
            NameError & 852 & 8.3\% \\
            SyntaxError & 753 & 7.3\% \\
            IOError & 167 & 1.6\% \\
            IndentationError & 153 & 1.5\% \\
            SystemExit & 115 & 1.1\% \\
            EOFError & 94 & 0.9\% \\
            OSError & 48 & 0.5\% \\
            ValueError & 34 & 0.3\% \\
        \end{tabular}
        \vspace{-0.5cm}
    \end{table}

     Table~\ref{tbl:evaluation-before-naive-inference} provides the names and counts for the most common reasons a gist terminated when run in an isolated Python \texttt{v2.7.13} environment.

     Consistent with the Yang et al.~\cite{Yang:2016} study on Stack Overflow Python snippets, we observed that only 24.4\% of Python gists were executable. The majority of gists (52.4\%) failed to execute due to an \texttt{ImportError}, which is typically caused when a python dependency could not resolved or loaded. We observed that only 17.1\% of gists failed to parse (i.e., \texttt{SyntaxError}, \texttt{NameError}, and \texttt{IndentationError}. Our observed rate of parse failures for gists is slightly lower when compared with Yang et al.'s observed rate of 25\% for Stack Overflow snippets. We believe this may be caused by the difficulty of distinguishing exposition from code when parsing code snippets found on Stack Overflow~\cite{Sillito:2012}. For example, in a Stack Overflow post, it could be common to include code and output typed into an interactive shell in order to help explain a concept, which is not directly parsable.

     Finally, we observed \textless 8\% of gists failed to execute due to some other runtime exceptions, such as \texttt{IOError} or \texttt{OSError}. These failures could be caused by missing resources, such as files, services, or platform specific dependencies.

    Baseline results for executing in a Python 3 environment show 3,907 instances of \texttt{SyntaxError}, compared to the 753 for Python 2. In addition, the number of gists which exited with \texttt{Success} dropped to 1,445. The number of gists which exited with \texttt{ModuleNotFoundError}, a direct subclass of \texttt{ImportError} in Python 3.6, was 3,353. While this shows a decrease from the 5,379 in Python 2, the large set of \texttt{SyntaxError} may shadow an undetermined set of gists which would also see an \texttt{ImportError}.

\vspace{0.5em}
\noindent \insight{Overall, we find that most gists are not executable in a default Python environment. Further, the exceptions raised when attempting to execute the gists suggests that an insufficiently configured environment is the primary cause.}




\subsection{\RQEnv}\label{subsec:rqenv-results}

    The baseline analysis for \RQExecShort{} showed the majority of Python gists require environment configuration. To determine if a simple algorithm is capable of resolving such errors, we applied our inference algorithm described in Section~\ref{sec:inference} to the 5,379 gists which failed due to \texttt{ImportError} using Python 2, attempting to install all third party imports with \texttt{pip} in both a Python 2.7.13 and Python 3.6.5 environment. Python 2 is used as a baseline for \texttt{ImportError} due to its lower frequency of \texttt{SyntaxError}.

    We analyzed each gist after attempting to install all inferred dependencies and recorded the exit status according to the same criteria used for answering \RQEnvShort{}. For Python 2, 2,488 gists exited due to a reason other than \texttt{ImportError}, a gain of approximately 46\%. Of these gists, 1,294 finished with \texttt{Success}. The remaining 1,194 finished with some error other than \texttt{ImportError}. When also considering Python 3, the number of gists which had become executable increased to 2,870. Overall, considering Python 3 resulted in an additional 428 gists becoming executable after inference when compared with only using Python 2.

    \vspace{0.5em}
    \noindent{\insight{
        While a naive approach can infer dependencies for some gists, it fails to do so in the majority case.
    }}

\section{Execution Failures}\label{sec:execution-failures}

To answer, \emph{\RQWNEnv}, we inspected the gists to better understand why they failed to execute, even after applying our naive algorithm. First, we focused on gists failing with \texttt{ImportError}, which was the most common failure status. Then, we also inspected gists which failed for other reasons, such as \texttt{IOError}. Finally, we characterize the effort reported by developers when manually creating Dockerfiles for the failing gists.

\input{tables/tbl-inspection.tex}

\subsection{Gists Failing with ImportError}

We report our findings in Table~\ref{tbl:inspection}. Overall, the 24 developers participating in this study were able to submit a response for 218 out of the 240 gists assigned to them as a group. The average number of Dockerfiles received from each developer was 9, with a minimum of 3 and a maximum of 10.

In addition to the failures reported in Table~\ref{tbl:inspection}, 24 gists were considered flaky. Inference of flaky gists may have failed due to network or memory issues. One developer reported needing to increase the memory Docker was configured to use in order to properly install dependencies for one such gist.

Collectively, the developers indicated that they were unsuccessful in creating a working Dockerfile for an additional 78 gists. The feedback we gathered for such gists showed that even developers familiar with environment configuration may be unable to correctly deduce the correct specification for an arbitrary snippet of code. One developer, after referring to an existing Dockerfile related to the gist they were working with, wrote

\begin{quote}\itshape
    I attempted to adapt the Dockerfile listed above to run this gist, but was never able to get it working; needless to say I would not have been able to do it without the Dockerfile listed either; I attempted various other ways to install the android sdk (apt-get, etc), all of which failed; constantly ran into 404 errors with apt-add-repository; got ``No space left on device'' error when running listed Dockerfile in a virtual machine; the Dockerfile built when running natively, but I could not find a way to use the ``monkeyrunner'' command, as this gist is supposed to be run with ``monkeyrunner'' and not ``python'' (from what I understand); a great deal of time spent trying futilely to get this to work.
\end{quote}

We now focus on a selection of distinct failure causes.

\textbf{Names.} The most common case, as stated in Section~\ref{subsec:rqenv-results}, is when a resource name does not match the name of the package it belongs to. Resolving this situation often required the developer to search the package index, test multiple packages, or query developer resources such as Stack Overflow.

For example, one gist relied on the module named \texttt{i3}, but the developer found they had to install the package \texttt{i3-py}, resulting in the following Dockerfile:

\begin{minted}[fontsize=\footnotesize,linenos,xleftmargin=0.5cm,breaklines,autogobble]{Dockerfile}
    FROM python:2.7.13
    ADD i3_focus_win.py /
    RUN pip install i3-py argparse
    CMD ["python", "/i3_focus_win.py"]
\end{minted}

\textbf{System dependencies.} Missing C libraries were also a common issue. Many Python dependencies serve as bindings into C libraries installed as a system dependency, and fail to compile on installation because the system dependency is not present. In some cases, a dependency failed to compile because the Python Docker image did not include C build tools, such as cmake, that they relied on.

One such gist made use of the Python \texttt{hunspell} package, a wrapper for the C program Hunspell. The developer found that before using pip to install \texttt{hunspell}, they needed to add \texttt{RUN apt-get install libhunspell-dev -y} to their Dockerfile.

\textbf{Custom environments.} In some cases, a dependency was distributed as part of a separate execution environment. For example, one developer reported that a gist relied on the \texttt{bpy} module that ships with Blender. After installing Blender and still seeing an \texttt{ImportError}, the developer discovered a Stack Overflow post saying \texttt{bpy} can only be imported when running in Blender's bundled Python interpreter.

\textbf{Unlisted packages.} Several gists depended on packages which were not available through the PyPI or Aptitude package managers by default. Such packages require being installed from a separate repo, such as an Aptitude Personal Package Archive (PPA) or directly from a git based public repo.

In one example, a user commented on the Gist that they had difficulty importing one of the modules, even though they had installed the correct package.

\begin{quote}\itshape
ScissorPush?

 from kivy.graphics import ScissorPush
 ImportError: cannot import name ScissorPush
\end{quote}

Resolving this issue required installing an unreleased version of python-kivy that needed to be installed from a PPA.

\begin{minted}
[
fontsize=\footnotesize,linenos,xleftmargin=0.5cm,breaklines
]
{Dockerfile}
FROM ubuntu:16.04
RUN apt-get update
RUN apt-get install -y software-properties-common python-software-properties
RUN add-apt-repository ppa:kivy-team/kivy-daily
RUN apt-get update
RUN apt-get install -y python-kivy
ADD snippet.py /snippet.py
CMD ["python", "/snippet.py"]
\end{minted}

\textbf{Deprecated packages.} In other cases, gists relied on packages that are no longer maintained and can no longer be installed. Common causes are not supporting SSL, which pip now requires, not fixing known bugs which prevent installation, or even an entire package no longer being provided for distribution.

For example, the Python Quartz package has an omission in the manifest that prevents the requirements file from shipping with the package source. The developer is aware of the issue, but has declared they will not create a patch.

\begin{quote}\itshape
    To fix this problem, I have to include requirements.txt in MANIFEST.in so that the file will be shipped with the sources.

    Unfortunately, I abandoned this project a while ago and I am currently working on a complete rewrite...
\end{quote}

Sometimes, a package is still actively maintained, but the gist relies on features from a version which had reached end-of-life and is no longer being distributed.

\textbf{Configuration settings.} Some gists require additional configuration files which are not provided with the gist itself. For example, it was common to read in secret keys and values from a non-existing app.config file in order to read a setting such as \texttt{TWITTER\_API\_KEY}. These configuration files are not preexisting dependencies which can be installed.

\textbf{Language version.} Python 3 has introduced several new modules, like \texttt{urllib.request}, that are not present in Python 2. Gists that rely on these modules must be run in a Python 3 environment, and are incompatible with the \texttt{python:2.7.13} Docker image being used. In some cases it may still be challenging to determine which Python version to use. For example, \texttt{pathlib} is a part of the Python 3 standard library, but was not introduced until Python 3.4, and support for it was only added to the standard library in Python 3.6.

\textbf{Operating System.} Developers also saw dependencies which could only be installed on a specific operating system, such as Windows or macOS. One developer, when asked to create a configuration for a gist, found that the gist was designed to interact with the Windows registry, and reported

\begin{quote}\itshape
    Packages are dependent on Windows (not Ubuntu).
\end{quote}

Such gists cannot be run in the Ubuntu based Python image.

\subsection{Other Failing Gists}




    To characterize the gists in our dataset and gain a better understanding of how they are used on GitHub, we computed basic metrics across all gists using tools developed for our execution harness. Additionally, we performed an inspection on 30 randomly selected gists from the 10.3k in our dataset with the focus on characterizing what resources they might rely on, including, but not limited to, dependencies.

    Our random sample found that 14 out of 30 gists (46\%) did not rely on a third party package. Approximately 13\% did not import any packages, and 76.7\% relied on Python library packages. 6.7\% optionally loaded a third party package if it existed in the environment.  We found that many gists rely on connecting to networked resources, or on interacting with configuration files and executables on the file system. Other gists required interaction from the user in some manner, either requiring input over \texttt{stdin}, command line arguments, creating an interactive prompt, or displaying information through a graphics interface. In the worst case, a gist does nothing because it is either recognizably not correct Python syntax, or because it defined classes or functions but did not otherwise execute code. This happened nearly 10\% of the time.

    Overall, the gists in our dataset import over 1,700 unique third party packages and on average have 92 lines of code.





\subsection{Developer Extraction Effort and Effectiveness}

The median difficulty rating reported for configuring a gist was 3 on a scale of 1-5, reported for 24.3\% of all gists. Only 13.7\% of the gists were reported as very easy to execute by our developers, whereas 22.4\% were reported as very difficult to execute. Developers reported spending between 20 minutes to 2 hours to setup the environment for executing each gist.

Of the 140 gists developers found an environment configuration for, the average Dockerfile was less than 10 lines and installed less than 5 packages. However, we found that not all of the submitted Dockerfiles were capable of executing their gists without \texttt{ImportError}. For example, one developer submitted the following Dockerfile, claiming that the gist ran without any errors in the provided environment.

\begin{minted}[fontsize=\footnotesize,linenos,xleftmargin=0.5cm,breaklines,breakafter=/,autogobble]{Dockerfile}
    FROM python:2.7.13
    ADD https://gist.githubusercontent.com/awesomebytes/cb5a28fa8d4db3fc1ba51894663c1aed/raw/cba597a5219d807c5e4940e9d2018d47b5eca809/watson_ros_publish_string  /snippet8.py
    RUN pip install ws4py
    CMD ["python","/snippet8.py"]
\end{minted}

However, we found that executing the gist still failed with the configuration error \texttt{ImportError: No module named rospy}. One interpretation is that not only can this be a time-consuming task for developers, but the process can be also error-prone.


\subsection{Developer Responses}

Section~\ref{sec:execution-failures} revealed common properties of gists that made environment configuration difficult. We now highlight a selection of developer responses which illuminate the process that developers employ when faced with these challenges.

\textbf{Version errors.} Developers reported several experiences related to resolving errors that were present due to mismatches in versions of dependencies and code.

    \begin{quote}\itshape
        django was the only import required. But that didnt simply resolve the error. There was a import error for CompatCookie. Tried in python 3 as well but no luck. Later found out from django release notes that it was deprecated after v1.4. So tried to pip install older version of django and was finally able to resolve the import error. Docker file builds and runs without any error.
    \end{quote}

Another developer described how the requirements could shift depending on the version of a dependency used.

    \begin{quote}\itshape
        Spent over an hour to find the imports needed for text.blob. It was replaced to textblob from version 0.7.1 and when I tried the lower version I received another error that required dependencies on a higher version.
    \end{quote}

\textbf{Unlisted or unknown dependencies.} Developers reported several instances where they had difficulty determining the provenance of a dependency.

    \begin{quote}\itshape
    1. Git clone basic-python-logger repo from https://github.com/vehrka/basic-python-logger (Have to google and find out that basiclogger.py is not a module, but rather a script wrote by the creator of the Gist itself) 2. pip install psycopg2, pandas, sqlalchemy for satisfying the dependencies. 3. Upon doing this, the images builds successfully
    \end{quote}

    Another developer had difficult working with a cloud provider package.

    \begin{quote}\itshape
         Couldn't find clouddns module. Couldn't solve dependency. Spent 2 hours on it.
    \end{quote}








\textbf{Resource limitations.} Several developers reported experiences related to memory or disk limitations on their personal computers when building environments.

    \begin{quote}\itshape
        MEMORY ERROR while installing keras and pyspark resolved by --no-cache-dir flag
    \end{quote}

\subsection{Summary}

We conclude \RQWNEnvShort{} with the following observation:

\vspace{0.5em}
\noindent{\insight{
    Python gists often require non-trivial environment configuration in order to run. There are multiple reasons why configuration for any particular gist might be difficult, but the most common challenges are finding dependencies without obvious names and installing dependencies with transitive dependence on system modules.
}}

\section{Discussion}

\subsection{Towards automated environment configuration and beyond}


While the inference procedures presented in \TOOL are simple, we show that they successfully lead to a correct environment specification in a number of cases. This indicates that reliable environment inference is possible, and highlights areas of research where techniques can improve. For example, we may consider combining dynamic inspection of packages and machine learning algorithms for inferring possible environment specifications.

Although we focus on Python gists, we believe our insights can generalize to other programming languages. For example, dependency on system build tools can also be a problem in the Node.js ecosystem when packages compile native addons. Common compilation troubles in Node eventually prompted Microsoft to publish developer guidelines~\footnote{https://github.com/Microsoft/nodejs-guidelines}.

Given a language which supports third-party packages, our approach only assumes two things. First, that packages have a set of named resources that they make available for use by client code. Second, that the identifier for a package resource used by client code has at least a substring match for a resource provided by the package. This is the case for popular languages like Javascript and Ruby, and so we believe that our approach will generalize to these, and similar, languages.

Furthermore, our inference procedure only requires a code snippet, and could easily be modified to work with another context, such as code snippets in Stack Overflow answers, blog posts~\cite{Parnin:blogs:2013}, or online documentation~\cite{Treude:2018}. Gistable focuses on configuring and running single file scripts. However, many projects have a large number of interacting tools that make configuration challenging. We believe insights from our work can inform configuration techniques for larger projects in the long-term.


\subsection{Challenges in mining gists}\label{sec:challenges}

\paragraph*{Querying for unique gists isn't directly possible} Instead, we rely on manipulation of search parameters in the GitHub UI to return results. On subsequent runs, the gists returned by a UI search are often different, allowing the use of a dictionary approach for collecting unique gists by ID. However, some gists may still be duplicates, either by forking or simple duplication of content. Forked gists have metadata available indicating the origin, but, in the worst case, it is generally undecidable if two gists are equivalent.

\paragraph*{Gists can be complex} While most of them are relatively simple, there is no requirement that a gist consist of only one file, or even of files in a single programming language. If a gist has more than one file, the entry point is often ambiguous, unless the programming language runtime supports running a default file, and such a file exists in the gist. We discarded gists with more than one file to avoid having to deal with this situation.

\subsection{Challenges in automated configuration inference}

There are several challenges identified by our work.

\paragraph*{Name resolution} An important task in automatically creating an environment specification from code is: \emph{given a code snippet, infer the set of installable packages associated with the code}. Luckily, package import statements within the code snippet can help; however, there are still several complications that must be resolved. In the simplest case, many package names may not match the name they are imported by (e.g. the \texttt{i3}/\texttt{i3-py} mismatch encountered by one of the developers participating in the study).

Another consideration is that many gists have imports structured as follows: \texttt{import kazoo.client}. In our evaluation, the naive algorithm attempts to install \texttt{kazoo.client}, and fails. The actual package is \texttt{kazoo}. However, in other cases, like \texttt{zope.interface}, the appropriate package name is indeed \texttt{zope.interface}. Finally, it could be possible that some code snippets are incomplete; that is, they may omit import statements for packages being used in code.

A first step to addressing this challenge may be to preprocess known packages by extracting a list of resources that each exports. When performing inference, resources might be mapped to installable packages by a reverse look up. However, this introduces its own challenge of dealing with packages which have conflicting resource names.

\paragraph*{System Dependencies} Other packages have implicit dependence on system environment configuration or other system packages. Unfortunately, this type of error often presents itself as a compile time error when a header file cannot be found. Header files, like package resources, do not necessarily have a name related to their project. Like the Python package name resolution challenge, this could be addressed by preprocessing and reverse look up. Another option is to analyze existing configuration scripts for Python projects and perform association rule mining to infer dependence between a Python package and a system dependency.

\paragraph*{Language Version} Even with a gist consisting of a single snippet in the desired language, it is often non-trivial to decide which language version to use. For our Python gists, most of them are capable of running successfully with Python 2. However, reported instances of \texttt{SyntaxError} may be due to use of syntax created in Python 3. Gists can be checked for syntax errors by attempting to compile them, so Python 2 or Python 3 syntax compatibility could be checked by compiling under each language runtime. Python 3 dependence may also be inferred by checking gist imports against the Python 3 standard libarary.

\paragraph*{Unlisted and Deprecated Packages} Packages may not be installable from a general package repository, like the Python package PyTorch. In this case, it may be possible to install directly from a git based repo, if one can be inferred from previously seen configuration scripts.

\subsection{Future Applications}
While the focus of our paper was evaluating the executability of Python gists, we
envision several additional research applications for \TOOL.

\begin{itemize}[leftmargin=*]
\item \textbf{Text summarization of tasks:} Recent work (\cite{Lucia:2014:LSC:2674501.2674546, Haiduc:2010:UAT:1919284.1919577, McBurney:2014:ITM:2597008.2597793, Eddy:2013}) has focused on performing semantic code summarization. Because gists typically correspond to idiomatic programming uses and tasks, there is an opportunity to use gists as a dataset for learning models which support semantic summarization of code.
\item \textbf{API usage analysis:} We observed over 1,700 unique third-party python packages in our initial version of \TOOL. This suggests that gists can provide a rich source of information for mining and understanding how APIs are used in practice by programmers.
\item \textbf{Test input generation:} Gists often have hard-coded input text for running the code example. An interesting research opportunity would be to use gists as a benchmark for generating test inputs that can also successfully run (or fail) in a gists.
\item \textbf{Configuration repair:} In addition to inference of configuration environments, it is possible to support research in repair of configuration scripts~\cite{Weiss:Tortoise:2017}. For example, if a user updates the code to use new library, pushes changes to production, but did not update the Dockerfile~\cite{Schermann:2018}, an exception can be thrown.  However, a configuration repair tool can suggest a repair that updates the environment specification to use the correct version of the package: e.g., \texttt{"networkx==2.0"}, which eliminates the exception.
\item \textbf{Resource repair:} Gists may have external resources, such as URLs, or publicly hosted APIs, such as the Google Maps API. One interesting application would be to study the decay of resources overtime (bitrot). Further, if resources or API URLs change, is it possible to repair the invalid resources and code?
\item \textbf{Understanding the Python community's use of gists:} Wang et al.~\cite{Wang:2015:GUU:2820518.2820556} studied the use of Public gists on GitHub, observing a variety of uses, but did not focus on usage categories per language or file type. We observed several different usage patterns for Python gists in the Gistable database. Future research can inspect and categorize the types of practices that emerge from creating and sharing public gists in the Python community.
\end{itemize}

\section{Limitations}

Our analysis may overestimate the number of executable snippets. For example, a gist may define only a single function containing all code, but never call it. So long as the function definition succeeds, the gist is marked as successful regardless of whether or not the code works. Any additional dependency errors caused by executing the function will not be triggered. Further, our analysis may misclassify an exception. For example, it is possible for a gist to  implicitly hide import errors by catching them and then raising an error of a different type. Future analysis can use several measures to increase the certainty of successfully execution: annotating gists with test assertions, increasing path coverage of executed gists, manual inspection and verification, and iteratively fixing configuration issues and evaluating gist execution.

Another concern is that running a gist once will only produce the first fatal error encountered, although the gist may have more than one. As a result, we may underestimate the distribution of some category of errors. However, research from Becker et al.~\cite{Becker:2018} argues that the practical approach in such a case is to focus on the first failure encountered, as this mirrors how developers typically resolve errors. 

We caution readers to not overgeneralize our results. While we analyzed a large sample of Python gists, these results may not extend to other programming languages. Numerous factors, such as the experience of programmers, the quality of modules and package management, the degree of third-party modules usage, and language design can influence how executable a code snippet is in practice. In other languages and ecosystems, these factors may be less of a concern. Further, we examine public gists, which may differ from private gists.

Our environment inference algorithm can have several limitations. Even if all dependencies install correctly and gist execution succeeds, there is no guarantee the package API will not undergo a breaking change between the time the Dockerfile is created and the time the image is built.



\section{Related Work}

The work by Yang et al.~\cite{Yang:2016} is the closest related work in terms of research approach and methodology. Yang et al.~\cite{Yang:2016} examined Python snippets on Stack Overflow and found that 75\% were parsable and only 25\% were runnable. In this paper, our work differs in several important ways. First, our primary focus is on the ability to execute Python snippets, whereas Yang et al. were primarily focused on the ability to parse snippets.
Second, we investigate the effectiveness of a naive inference algorithm in recovering an execution environment for the snippets. Third, we manually construct execution environments and characterize reasons why code may not be executable. Finally, our research context differs in that we examine gists shared on GitHub instead of code snippets found in Stack Overflow answers. Overall, our research complements Yang et al.'s~\cite{Yang:2016} work in understanding challenges for sharing and using snippets on the web, while providing new directions for research in automated configuration inference.

Several researchers have characterized the buildability of software projects.
Sulír and Porubän~\cite{Sulir:2016} performed a study on 7,200 Java projects and studied the ability to automatically build them by attempting a maven or ant build in a virtual machine. They found that more than 38\% of builds ended in failure. The authors identified that the largest portion of errors are dependency-related. Incidentally, Urli et al.'s study~\cite{Urli2018} on program repair of Java programs is related to our work. Urli et al. found that by attempting to automatically build 1,609 Java projects on GitHub with Maven, they could only reliably reproduce 31.82\% of test failures due to the complexity of mimicking configuration for test environments. A notable difference is that our work focuses on automated configuration inference, whereas Urli et al. focus on repairing Java code in order to pass test failures and thus does not investigate why projects could not build or run tests. \emph{Buildability} and \emph{executability} are related yet distinct concerns in software maintenance. First, build failures can be associated with difficulty inherent in build maintenance that is independent of reproduction. For instance, McIntosh et al.~\cite{McIntosh:2011} find that the effort involved in maintaining the build configuration can introduce 27\% overhead on source code development and a 44\% overhead on test development. Such high effort could increase the odds of out-of-date or non-buildable projects. Second, while building large and complex projects can be daunting, this process does not necessarily \emph{run} the code, which can require further environment specifications. Finally, our research context differs from buildability of software projects in that we are interested in automatically executing isolated code snippets without build specifications, which is common in learning or documentation contexts.

German et al.~\cite{German:2007} describe multiple problems associated with managing and specifying dependencies, including downloading, building, and satisfying inter-dependent artifacts, which may not always be explicitly documented.  They propose a framework for categorizing dependency types and a method for building and visualizing an inter-dependency graph of a package. Lungu and colleagues~\cite{Lungu:2010} note that dependencies also exist between projects in a software ecosystem. They propose a model which can capture inter-project dependencies. In our work, we are interested in characterizing both dependencies as well as other environment resources that when absent can prevent code from being executable. We believe our empirical findings complement these models and together, they can be used to inform the design of an automated configuration inference tool.


\section{Conclusion}

Code snippets can be a useful way to explain and demonstrate  a  programming  concept, but may not always be directly executable.
We investigated the executability of Python gists hosted on GitHub and the ability for a naive inference algorithm to recover a Dockerfile capable of executing the Python gist. Finally, we investigated the types of execution failures encountered when running Python gists and the effort involved in manually creating a Dockerfile able to run a gist.


Overall,  we  find  that  most  gists  are  not  executable  in
a  default  Python  environment.  Further,  the  exceptions
raised when attempting to execute the gists suggests that
an  insufficiently  configured  environment  is  the  primary
cause.

Our inference algorithm shows that, at least in some cases, correct application environment configurations can be automatically recovered. While a naive approach can infer dependencies for some
gists, it fails to do so in the majority case. Additional strategies promise greater success, and will be the subject of future research.

Our investigation of Python gists finds that they often require non-trivial environment configuration in order to run. There are multiple reasons why configuration for any particular gist might be difficult, but the most common challenges are finding dependencies without obvious names and installing dependencies with transitive dependence on system modules.

Finally, we envision multiple applications for \TOOL that extend beyond empirical studies of executability. \TOOL can automatically configure and execute approximately 5,000 public Python gists hosted on GitHub. Each gist has an accompanying Dockerfile which can be used to build a Docker image based off of the \texttt{python:2.7.13} image which contains both the gist and its dependencies. Running the Docker image executes the gist without \texttt{ImportError}. \TOOL also ships with a simple command line utility for cloning gists in the dataset, and building and running Docker images.



\bibliographystyle{IEEETran}
\bibliography{references}

\end{document}

%% file: macros.tex
\newcommand{\TOOL}{Gistable\xspace}

\newcommand{\RQExecShort}{RQ1}
\newcommand{\RQExec}{\RQExecShort{} -- Can gists be executed?}

\newcommand{\RQEnvShort}{RQ2}
\newcommand{\RQEnv}{\RQEnvShort{} -- Can a naive algorithm enable exectuable gists?}
\newcommand{\RQWNEnvShort}{RQ3}
\newcommand{\RQWNEnv}{\RQWNEnvShort{} -- Why might gists not be executable?}

\newcommand\insight[1]{
	\noindent 
	\fcolorbox{gray!20}{gray!20}{
		\parbox{0.92\columnwidth}
		{#1}
		\hspace*{0.5ex}
	}
}

%% file: tables/tbl-inspection.tex
\begin{table*}
    \centering
    \begin{threeparttable}
        \caption{
            We had 24 developers familiar with environment configuration techniques attempt to
            manually create Dockerfiles for 218 of the gists for which naive inference failed to resolve
            import errors. This table summarizes reasons for failure as reported by the developers, focusing
            on the first failure reported. We manually inspected each gist in cases where no clear reason
            was found by a developer, applying our own failure category if possible, or labeling the gist
            as unconfirmed.
        }
        \label{tbl:inspection}
        \begin{tabulary}{\textwidth}{p{0.5\textwidth}p{0.05\textwidth}p{0.35\textwidth}}
            \toprule
            \textbf{Cause} & \textbf{Count}  & \textbf{Example} \\ 
            \midrule
            
            
            Package name did not match the resource imported in the gist &
                70 & \url{https://gist.github.com/syl20bnr/6623972} \\ \midrule
            Gist dependencies have additional dependencies which need to be resolved &
                23 & \url{https://gist.github.com/kennethreitz/2901479} \\ \midrule
            Relies on missing C library files or headers & 
                16 & \url{https://gist.github.com/huyx/8069261} \\ \midrule
            Requires a previous version of a package due to breaking changes &
                15 & \url{https://gist.github.com/segphault/9f2d7da68779a17a0890} \\ \midrule
            Dependency can only be installed on a non-linux operating system &
                13 & \url{https://gist.github.com/mapleray/4189391} \\ \midrule
            Relies on a standard package that was introduced in a later version & 
                12 & \url{https://gist.github.com/fmasanori/4684752} \\ \midrule
            Pip errored during installation, possibly timing out on large packages or propagating an exception raised by the package &
                12 & \url{https://gist.github.com/willwade/5330566} \\ \midrule
            Unconfirmed. The exact failure could not be narrowed down to a single category. &
                9 & \url{} \\ \midrule
            Gist is missing necessary environment configuration, such as settings files &
                8 & \url{https://gist.github.com/Sinkler/bfc2099235ac96937f34} \\ \midrule
            Dependency wasn't available on PyPI, nor installable via the Ubuntu aptitude package manager. &
                7 & \url{https://gist.github.com/JudoWill/764262} \\ \midrule
            Dependency is only supplied as part of a custom execution environment or interpreter & 
                6 & \url{https://gist.github.com/Utopiah/a2b9c6ecdb24ca8fd6f4f41a9c0eb32e} \\ \midrule
            Relies on a deprecated package that is no longer maintained and is no longer available to be installed & 
                1 & \url{https://gist.github.com/matbor/6532185} \\ \midrule    
            Gist is not intended to be run and imports libraries which don't exist &
                1 & \url{https://gist.github.com/RichardBronosky/454964087739a449da04} \\ \midrule
            No versions are available for install &
                1 & \url{https://gist.github.com/mclavan/276a2b26cab5bc22d882} \\
                
            \bottomrule
        \end{tabulary}
    \end{threeparttable}
    \vspace{-0.5cm}
\end{table*}